\documentclass[10pt]{article}
\usepackage{amssymb}
\usepackage{amsthm}
\usepackage{amsmath}
\usepackage{float}

\usepackage[pdftex]{color,graphicx}

\usepackage{hyperref}
\hypersetup{colorlinks=true,allcolors=blue}
\usepackage{hypcap}
\usepackage{setspace}

\title{Market Impact Paradoxes
}
\author{ Igor Skachkov\\iskachkov@yahoo.com }

\begin{document}
\maketitle
\begin{abstract}

The market impact $(MI)$ of Volume Weighted Average Price $(VWAP)$ orders is a convex function of a trading rate, but  most empirical estimates of transaction cost are concave functions. How is this possible?
We show that isochronic (constant trading time) $MI$ is slightly convex, and isochoric (constant trading volume) $MI$ is concave.
We suggest a model that fits all trading regimes and guarantees no-dynamic-arbitrage.
\end{abstract}

\pagebreak
\tableofcontents
\clearpage
\newpage
\section{Introduction}

In a recent paper by Farmer et al. \cite{FGLW2013} $(FGLW)$ the authors name three reasons for studying market impact: theoretical (market impact shape reflects general laws of economics), ecological (market impact makes large fund managers diversify their assets) and practical (correct evaluation of market impact is essential for optimal trading strategies).

In this paper we focus almost entirely on the practical aspect of market impact theory. More precisely we try to find the functional form of market resilience to large parent order execution.\footnote{I am used to the term \emph{parent order} for a large order that should be executed during the current trading session or in the next few days and the term \emph{child order} for the fraction of the \emph{parent order} that would be directly submitted to the exchange. I suspect that these terms came from developers of trading applications and were inherited from C++/Java languages syntaxis. Another term \emph{hidden order} is a descriptive term from the point of view of high  frequency traders who try to detect those large orders. The term \emph{hidden order} is also applied to the exchange orders not displayed to the market. $(FGLW)$ used the term $metaorder$ and I will use this name interchangeably with \emph{parent order} }. Unfortunately empirical data are controversial, they are aggregated and filtered under different and not transparent conditions and sometimes are not consistently interpreted. In this section we give short description of the market impact kernels that allow close form analytical solutions for optimal trading trajectories. In  section \ref{S:Diff} we present a diffusion kernel and in the next sections we show how this model can help to resolve the paradoxes between market impact theory and empirical data.
%
%

\paragraph{Analytical market impact models}

In short time horizon we assume the price $S$ dynamics:
\begin{equation}
S=S_0+h[x]+S_0 \sigma W_t
\end{equation}
where $h[x]$ is a temporary market impact functional that in general depends on execution history and $S$ is the mid-price, i.e. the average price of Best Bid and Offer $(BBO)$ quotes.

\begin{equation}
\label{E:temp market impact}
h[x]=\int^t_0{f(q(\tau)) K(t-\tau)}dt
\end{equation}
Market impact kernel $K(\tau,t)$ is assumed to be a convex monotonic decreasing function, homogenous in time $K(\tau,t)=K(t-\tau)$ .

\subparagraph{GKAC model}
At the end of the last century  Grinold and Kahn $(1999)$ \cite{GK2000} and  Almgren and Chriss $(1999)$ \cite{AC1999,AC2001,AC2003} $(GKAC)$ independently pioneered application of calculus of variation to the problem of portfolio liquidation. Today most modern trading engines  use different modifications of their method.

They suggested mean-variance utility
\[
\Phi=\int^t_0{(E(R)-\tilde{\lambda} Var(R)})dt
\]
that is given by the functional
\begin{equation}
\label{E:utility}
\Phi [x]=\int^T_0{(h[x]\dot{x}-\tilde{\lambda} (S x \sigma)^2 })dt
\end{equation}
Calculating variations
\footnote{
\[
\begin{split}
&\delta_{\dot{x}}\Phi=\int^T_0{[\delta (\dot{x}(t)) \int^t_0{\dot{x}(\tau)K(t-\tau)d\tau}+\dot{x}(t)
\int^t_0{\delta( \dot{x}(\tau))K(t-\tau)d\tau}]dt}\\
& \text {We change the order of integration for the second integral and get}\\
&\delta_{\dot{x}}\Phi=\int^T_0{ \int^T_0{\dot{x}(\tau)K(|t-\tau|)}d\tau \delta (\dot{x}(t))dt}\\
\end{split}
\]
}
we obtain the following equation
\begin{equation}
\label{E:kernel euler}
2 k^2 x=  \frac{d}{dt}\int^T_0{\dot{x}(\tau) K(|t-\tau|)d\tau}
\end{equation}

$R$ is an absolute asset price return in \$

$Var$ is a variance

$T$ is a time horizon.

$X_0$ and $X_T$ are initial and terminal position in a stock

$x$ is the current position

$\dot{x}$ is its time derivative

trading rate $q=-\dot{x}$ is positive when cash flow goes in

$\eta$ is a coefficient of temporary market impact

$\tilde{\eta}=\eta \frac{\sigma }{A \! D \! V S_0}$

$A \! D \! V $ is an \emph{Average Daily Volume}

$\tilde{\lambda}$ is a risk-averse parameter.

$\lambda=k^2=\tilde{\lambda} / \tilde{\eta}$

$f(q)$ is reduced to a linear function of trading rate, $f=-\tilde{\eta}\dot{x} $

The simplest form of the convolution integral kernel was proposed
\[
K(t)=\delta(t), \qquad Dirac's \ delta \ function
\]

and a nice analytical solution obtained

\begin{equation}
\label{E:AC_solution}
x\left(t\right)=X_0 \frac{\sinh\left(k(T-t)\right)}{\sinh\left(kT\right)}+X_T \frac{\sinh\left(kt\right)}{\sinh\left(kT\right)}
\end{equation}

\subparagraph {Exponential kernel}
The solution (\ref{E:AC_solution}) is an acceptable approximation for optimal trajectories in practice, but it cannot describe the market impact of a single discrete trade (it is a delta-function). The instantaneous recovery assumption is unrealistic and inconsistent with calibration procedures. To resolve these problems one has to replace the delta function with a smooth kernel. The next step after Dirac's delta function is the exponential kernel

\[
K(t) \sim exp(-\beta t)
\]

The general solution for the optimal trajectories would be given by

\begin{equation}
\label{E:general solution exp kernel}
x=C_1e^{kt}+C_2e^{-kt}+D_1\mathcal{H}(t)+D_2\mathcal{H}(t-T)
\end{equation}

where $\mathcal{H}(t)$ is a Heavyside's function.
We need four equations to find four arbitrary constants. Two equations represent initial and terminal conditions and two additional equations follow from the requirement that $x(t)$ doesn't have $exp(\pm \beta t)$ terms. After some simple but tedious algebra we get a solution in a familiar form:

\begin{equation}
\label{E:nice_exp kernel solution}
\begin{split}
&x\left(t\right)=X_0B\frac{\sinh\left((k(T-t)+A\right)}{\sinh\left(kT+2A\right)}+
X_TB\frac{\sinh\left(kt\right)}{\sinh\left(kT+2A\right)}\\
&where\\
&A= \ln{\sqrt{\frac{\beta+k}{\beta-k}}}, \quad B=\frac{k}{\sqrt{\lambda}} , \quad k^2=\frac{\lambda \beta^2}{\lambda + \beta^2}
\end{split}
\end{equation}

More detailed derivation and more general solution can be found in Skachkov \cite{iS2010}.
For risk-neutral traders $(\lambda\rightarrow 0)$ the optimal schedule under exponential impact relaxation is a combination of two jumps and straight line between them.

\begin{equation}
\label{E:exp kernel lambda 0}
\begin{split}
&\lim_{\lambda \to 0} x=(X_0-\Delta X_0)\frac{T-t}{T}+(X_T+\Delta X_T)\frac{t}{T}\\
&\Delta X_0=\Delta X_T=\frac{X_0-X_T}{\beta T+2}
\end{split}
\end{equation}

Optimal trading strategy with the exponential kernel was the subject of Obizhaeva and Wang study \cite{OW2005}. They were the first to point out that discontinuity of optimal paths at the ends of a time interval and to derive optimal risk-neutral trajectories (\ref{E:exp kernel lambda 0}).
With $\beta \to \infty$ our result (\ref{E:nice_exp kernel solution}) goes to the classic solution (\ref{E:AC_solution}).

\pagebreak

\section{Market impact as a diffusion process }\label{S:Diff}
\begin{figure}[!hbt]
\begin{minipage}[b]{0.45\linewidth} 
\centering
This is a scheme of a drill stem test.
\resizebox{1.0\textwidth}{!}{\includegraphics{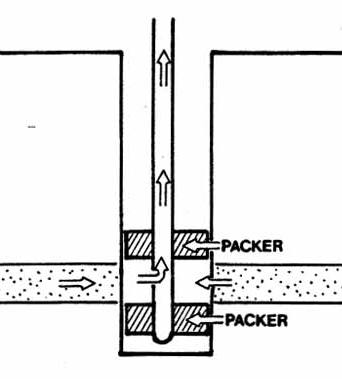} }
\caption{\footnotesize A drill stem test.}
\label{fig:dst}
\end{minipage}
\mbox{\hspace{0.5cm}} 
\begin{minipage}[b]{0.45\linewidth}
\centering
\resizebox{1.0\textwidth}{!}{\includegraphics{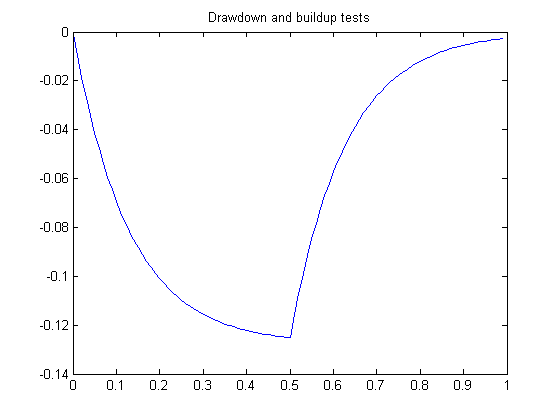} }
\caption{\footnotesize drawdown and buildup tests. The well worked with the constant rate, then it was closed }
\label{fig:ddandbuildup}
\end{minipage}
\end{figure}

Two models for the kernel of convolution integral are natural first choices due to their simplicity. In both cases we have analytical solution for optimal trajectory. The $GKAC$ model is memoryless and it means that the history of trading is unimportant. With exponential decay (market perturbations decay the same way as radioactivity) we also don't need to know the full history  - only the current state matters. This is a great relief for the developers of trading engines. If the program was interrupted, it can be restarted from scratch with new values for stock positions or with a single additional parameter - the difference between expected current price and expected equilibrium price. Both models proved their usefulness for trade scheduling in a continuous regime.
Unfortunately both of them contradict the empirical evidence of market long memory. It is almost generally accepted that market perturbations decay as a power law and that the square root is a good approximation in practice. Another desirable property is the absence of \emph{price manipulation strategies}, i.e. the possibility to make money on a \emph{round trip} $(X_0=X_T)$ execution. This regularity condition was introduced by Huberman and Stanzl \cite{HS2004} and analyzed in details by Gatheral \cite{G2010} and, in a later review, by Gatheral and Schied \cite{GS2013}. Some models do not admit price manipulation but their optimal execution trajectories may oscillate strongly between buy and sell trades \cite{AS2012}. This is a common problem with the fitting of real natural or social phenomena by abstract functions. 
To avoid the complications of odd effects of arbitrary general functions we borrow a model under which good behavior is guaranteed. A promising candidate for the decay kernel is the solution of a diffusion equation. The transmission of pressure in a porous medium filled by slightly compressible fluid is governed by a system of linear parabolic equation and initial and boundary conditions:

\begin{equation}
\begin{split}
f_t - \kappa f_{xx}=0, \ 0<x<x_2 \\
f(x,t)=h(t), \ -x_1<x<0
\end{split}
\end{equation}
where $\kappa$ is a diffusion coefficient, and subscripts $t$ and $x$ denote partial derivatives with respect to time and space variables.
\begin{equation}
\label{E:sytemTX}
\begin{split}
f(0,x)=0\\
h(t)=f(t,0)  \\
c\cdot h_t=q(t) + \kappa f_x, \ x=0
\end{split}
\end{equation}
Initially the distribution is uniform.
The inner boundary conditions are: the first equation is the continuity of pressure (price) and the second is the balance of cash (liquid) flow and wellbore storage, $q(t)$ is a known trading (flow) rate and $Q(t)$ is a cumulative amount traded, $q=Q_t$. We assumed that the flow is a linear function of price gradient - that is Darcy's law in a porous medium fluid mechanics or Fourier's law in heat transfer.

The outer boundary condition at $ x=x_2\leq \infty $ allows to model permanent impact of trading flow to the equilibrium price in market (reservoir) with finite capacity.
\begin{equation}
f_x = 0 \ \text{\ (impenetrable wall at $x=x_2$)}, \   \label{Eq:inf}
\end{equation}
The solution of the system (\ref{E:sytemTX}) in Laplace domain
\begin{equation}
\label{E:solLaplX2}
\bar{h}= \frac{\bar{q}}{cs+\kappa\sqrt{\frac{s}{\kappa}}\tanh(\sqrt{\frac{s}{\kappa}x_2} )} = \bar{q}\bar{K}
\end{equation}
where $\bar{y}$ denotes Laplace transform: $\bar{y}=\mathcal{L}y$.
We focus on temporary impact in this paper and don't consider very long trading times. With the assumption that the outer border of our reservoir is much greater than the radius of investigation
\[
\sqrt{\frac{s}{\kappa}}x_2\gg 1
\]
the solution (\ref{E:solLaplX2}) can be simplified and analytically inverted back to time domain
\begin{equation}
\label{E:solLaplInf}
\bar{K}= \frac{1}{cs+\sqrt{s\kappa}}
\end{equation}
\begin{equation}
\label{E:solTime}
h_\delta(t)= K(t) = c^{-1}\cdot exp(\tilde{t}) \cdot erfc(\sqrt{ \tilde{t} }),
\end{equation}

where $erfc(t)$ is a complimentary error integral $erfc(t)=1-erf(t)$.
\[
\tilde{t} = \frac{\kappa}{c^2}t
\]

Immediately  after the shock
\begin{equation}
\label{E:solTimeSmall}
K(0)=c^{-1}
\end{equation}

Then perturbation decays as an inverse square root of time
\begin{equation}
\label{E:solTimeBig}
K(t)\simeq \frac{1}{\sqrt{\pi \kappa t}}, \ x{_2}{^2}\gg \tilde{t}\gg 1,
\end{equation}
to the permanent value
\begin{equation}
\label{E:solperm}
K(\infty)=\frac{1}{c+x_2}
\end{equation}

The market impact of a constant unit rate of trade $q(t)\equiv 1$ is an integral of Green's function or source function.  In Laplace domain the integration is just division by the Laplace variable $s$. As a free gift from the integral transform approach we can invert the equation (\ref{E:solLaplX2}) for market impact with the given trading rate and get the required rate that would sustain predefined price value. For example, if we want to keep the constant price difference with arrival price $\Delta S $, we need after initial shot ($q$ is Dirac delta function at the initial time) trade with the decreasing as square root of time rate.
\begin{equation}
\label{E:constprice}
q=c\delta(t)+\sqrt{\frac{ \kappa}{\pi t}}
\end{equation}
Correspondingly, to sustain $S \propto t^\alpha$ price growth, trading rate 
\begin{equation}
\label{E:powerPrice}
q=c \cdot \alpha t^{\alpha-1} + \sqrt{ \kappa} t^{\alpha-\frac{1}{2}} \frac{\Gamma(\alpha+1)}{\Gamma(\alpha+\frac{1}{2})}
\end{equation}
is required.

With the choice of diffusion process as a base for market dynamics we are guaranteed from surprises like oscillating optimal trajectories and dynamic arbitrage \cite{G2010}, \cite{GS2013}. The transaction cost in terms of porous hydrodynamics is the work (wealth) that needed to do for the process:
\begin{equation}
\label{E:work}
\Delta W=\int^{V(T)}_{V_0}{PdV(t)}
\end{equation}
where $P$ is a pressure (expected stock price), and $V(t)$ is a volume (current volume traded). The process is adiabatic, i.e. there is no exchange of heat (information) between a system and its environment. According to the first law of thermodynamics
\begin{equation}
\label{E:work}
\Delta W=\oint {PdV(t)} \geq 0
\end{equation}
Replacing `thermo' by `mercato'
\footnote{Latin word \emph{mercatus} - market + a
Greek word $\delta\upsilon\nu\alpha\mu\iota\varsigma$ (dynamis)- power. The name Agorodynamics (Greek $\alpha\gamma o\rho\alpha$ (agora) - market) is more consistent etymologically  }
we can postulate:

\emph{For a mercatodynamics cycle, the wealth supplied to a closed system, minus that removed from it, equals the net payment made by the system. It is not possible to construct a perpetuum mobile machine which will continuously trade without consuming wealth.}


Diffusion model is rich and elaborated, has a huge library of the problems in physics and engineering, that were solved and analyzed in details: namely in classic theory of conduction of heat in solids \cite{KJ1959} and in porous media hydrodynamics and modern oil and gas well test engineering \cite{Ho1995}.
For example, we can consider `skin effect' at the inner border of our `reservoir' or dual porosity (fractures - matrix) formation to model various deviation from standard behavior. Nonlinear problems of real gas pseudo-pressure dynamics are also well developed. Figures \ref{fig:dst} and \ref{fig:ddandbuildup} show the drill stem test \emph{(DST)} and plot pressure versus time during the drawdown and buildup periods of \emph{DST}.
Additionally, we are free to choose our space dimension $D$: arbitrary dimension diffusion equations lead to the same type modified Bessel equations. 
Asymptotic behavior for the stock price  that is being traded with the constant rate $q$
\begin{equation}
\label{E:asymptNDim}
h(t) \sim q
\begin{cases}
t^{1-D/2}, \quad &D<2 \\
ln(t), \quad &D=2
\end{cases}
\end{equation}
The inverse Laplace transform of equation \ref{E:solLaplX2} for market impact decay is the main technical result of this paper.

\begin{equation}
\label{E:GreenTimeDiff}
K(t)= \mathcal{L}^{-1} \frac{1}{cs+\kappa\sqrt{\frac{s}{\kappa}}\tanh(\sqrt{\frac{s}{\kappa}x_2} )}
\end{equation}
There is no practical need for finding any analytical simplifications in time domain, because the modern numerical inverse Laplace transform algorithms are fast and accurate (see references in \cite{iS2002}) for such  a smooth function as a diffusion equation solution.
Of course, similar, but not the same results can be obtained by choosing the arbitrary power function
   
\begin{equation}
\label{E:GreenTimePower}
K(t)= C_0 + \frac{C_1}{(t_0+t)^\alpha} 
\end{equation}

At this point we can consider the well-reservoir system as an augmented space and space dimension $x$ as an auxiliary dimension that allowed us to get a solution free of dynamic arbitrage and oscillations by construction. In the next section, we find the physical meaning of the coefficients of diffusion and wellbore storage and try to interpret space variable $x$.

\pagebreak

\section{Paradoxes in Market Impact Theory}
We derived a comparatively simple linear model that is dynamic arbitrage proof and correctly describes decay of market perturbations. In this section we are going to reconcile it with the numerous empirical evidences that market impact is a concave function of a trading rate for large parent orders and that \emph{VWAP} algorithm performance is flat if market participation is small $(<1\%)$ and then slightly convex \cite{SB2005} up to $50\%$ $ADV$. The good news is that those facts contradict each other.


\paragraph{Instantaneous market impact}

While market impact and price impact sometimes considered synonymous\footnote{A comprehensive introduction to the modern state of a market microstructure theory and a survey of recent publications was presented by Gould et. al \cite{GP2012}}, this is obviously not the case for instantaneous or impulse impacts. Mean relative depth profiles of Limit Order Book $(LOB)$ exhibit a hump shape in a wide range of markets, including the Paris Bourse, NASDAQ, the Stockholm Stock Exchange , and the Shenzhen Stock Exchange. The maximal mean depth available for SPY was reported to occur at best bid and best ask levels, which could also be considered as a hump with its maximum at a price of Best Bid and Offer $(BBO)$ \cite{GP2012}. This means that `unbiased` \emph{instantaneous price impact}  should have a complex shape: first concave, then convex.
We exclude the convex instantaneous impact as an almost purely theoretical artifact. In reality, this scenario only occurs when an accidently wrong big order with a wrong limit (nobody places pure market orders) sweeps a few best bid/ask levels of the book. Almost surely it is the human error of a programmer or a manual trader (or an indication of price manipulation). In most cases impulse price impact is equal to zero, otherwise it is equal to $1$ tick for liquid stocks \cite{Ge2007}.
In contrast, \emph{instantaneous market impact} is never zero in response to nonzero perturbation of $LOB$ and it is non-measurable directly. For example, we want to buy $200$ shares of $XYZ$ and see $1000$ on both sides of $BBO$. We submit either limit order on bid price or market order on ask. In both cases mid price is not changed and in both cases we expect that in some future time the market for $XYZ$ would be a bit higher. We don't know in advance the actual sizes of $BBO$, we don't know if hidden orders exist inside the spread - all we know is that  \emph{immediate}  mid-price is the same and \emph{immediate} micro-mid-price (mid-quote-price weighted by bid and ask sizes) changed proportionally to the size of our order.

The empirical studies by Cont, Kukanov and Stoikov \cite{CK2012} support that simple heuristic.
They used one calendar month (April, 2010) of trades and quotes data $(TAQ)$ for a set of $50$ $S\&P500$ stocks and introduced \emph{order flow imbalance}, a variable that cumulates the sizes of order book events, treating the contributions of market, limit and cancel orders equally, and provided  evidence for a linear relation between high-frequency price changes (from $50$ milliseconds to $10$ seconds) and order flow imbalance for individual stocks.
Extracting concrete numbers from market data is a job that never guarantees conclusive results: the samples cannot be representative without access to proprietary information, data is not clean enough and some filtration with additional assumptions is needed, e.g. for relaxation time. Data preprocessing is never neutral: it filters out `biased' (actually conditioned to intelligence of the engine and a trader who monitors it) and favors mechanical style execution. For example, $WVAP$ algorithm should stop execution if experiences severe adverse selection and sharp unfavorable price jumps - and this is the result of a good engine, not  bad data.


There is another and better way to evaluate instantaneous impact: to take average execution performance in  of $VWAP$ orders with low participation rate, say $0.1-2\%$ for liquid stocks.  The numbers would be close in mean-variance sense across large brokerages. Information about real trading engines is highly proprietary. Each institution has `unique' set of algorithms usually with fancy names and promises the best in the world and flexible execution. And they actually can be very different in some internal details. For example, Toth et.al \cite{TE2011} demonstrated the heterogeneity of \emph{London Stock Exchange (LSE)}  broker liquidity provision. Some use limit orders almost exclusively, others predominately use market orders.\footnote{Actually the first approach is impossible with high market participation} However, those details don't cause dramatic differences in engine performances: all major brokers have to provide cost-variance quasi-efficient order execution to clients.
The performances of major trading engines are similar regardless of order type preferences, and this is a proof that market and  limit orders with the high probability to be executed have similar impacts. For an oversimplified short description of a trading engine see  Appendix \ref{A:B}.

The engines usually generate the orders of typical for that particular stock size. Randomization of schedule and sizes can help to hide the intentions of directional traders from high frequency predators, but the picture for small and moderate in market participation transactions remains the same\footnote{
Consider  the \emph{(VWAP)} algorithm, that places the orders each $1000th$ trade. The shock that the market experienced after execution of our order probably would be mostly forgotten after $999$ similar shocks of different signs. The same would be probably true for $1\%$ market participation. It means that the response for the small infrequent perturbations is almost constant while the interactions between them are negligible.}. We posit

\newtheorem{corollary}{Corrollary}
\newtheorem*{theorem}{Efficient Trading Hypothesis (ETH)}

\begin{theorem}
Algorithmic trading is efficient.
\end{theorem}

\begin{corollary}
Parent orders are optimally split into child orders.
\end{corollary}
\begin{corollary}
The implementation shortfall has the minimum: it is the best transaction cost per share that can be achieved in one sided trading.
\end{corollary}

Continuous trading approximation doesn't work for low market participation $VWAP$ algorithm. Instead of
\begin{equation}
\Delta W= \frac{1}{Q} \int_{0}^{T} q(t) \int_{0}^{t} q(\tau)K(t-\tau) d\tau d t
\end{equation}
we have to evaluate discrete equation
\begin{equation}
\Delta W= \frac{1}{2Q} \sum_{j=1}^{N} \sum_{j=1}^{N} q_j q_i K(|t_j-t_i|)
\end{equation}
If we assume equal child orders $q_i=\tilde{q}$ and large enough intervals between their submissions $\tilde{t}_{i+1}-\tilde{t}_{i}\gg 1$,
\begin{equation}
\Delta W= \frac{1}{2} (\tilde{q}K(0)+QK(\infty)) , \ Q \ll ADV
\end{equation}

Figures \ref{fig:8tradesT} and \ref{fig:logImpactT} clearly show this situation:
\begin{figure}[!hbt]
\begin{minipage}[b]{0.45\linewidth} 
\centering
\resizebox{1.0\textwidth}{!}{\includegraphics{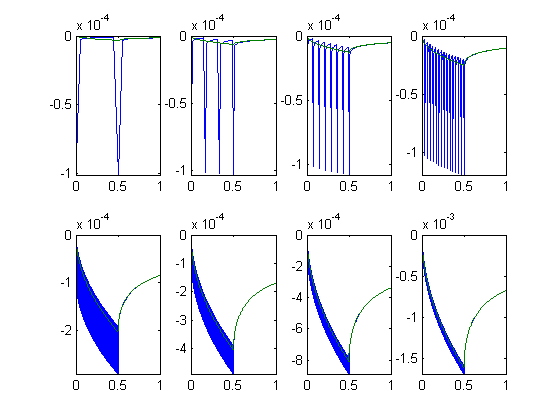} }
\caption{\footnotesize Market impact due to discrete and continuous trading. Isochronic regime, i.e. trading time $T$ is constant $T=0.5$ days. Single order size $Q_n=1e^{-4}ADV $. Number of trades: $1,2,4 \ldots, 1024$. Shown $2-16$ and $128-1024$ cases. }
\label{fig:8tradesT}
\end{minipage}
\mbox{\hspace{0.5cm}} 
\begin{minipage}[b]{0.45\linewidth}
\centering
\resizebox{1.0\textwidth}{!}{\includegraphics{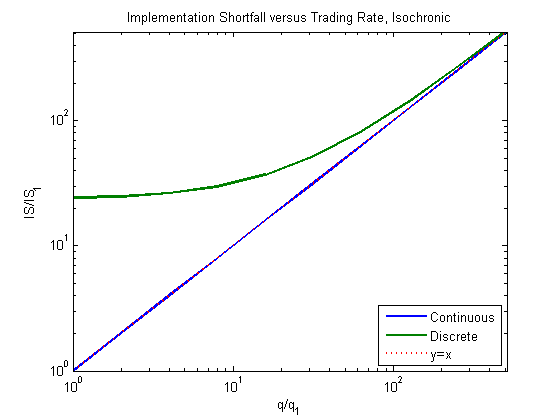} }
\caption{\footnotesize Loglog plot for trading cost versus trading rate from a Figure at left. Discrete (green line) and continuous (blue line) trading}
\label{fig:logImpactT}
\end{minipage}
\end{figure}\begin{figure}[!hbt]
\begin{minipage}[b]{0.45\linewidth} 
\centering
\resizebox{1.0\textwidth}{!}{\includegraphics{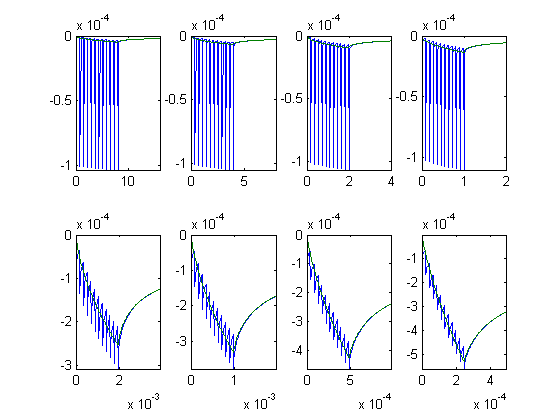} }
\caption{\footnotesize Market impact due to discrete and continuous trading. Isochoric regime, i.e trading volume $Q=12\cdot 1e^{-4}ADV$ is constant. First plot time horizon is 8 days. Then we divide time by 2 16 times. Shown first and last $4$ simulations.}
\label{fig:8tradesQ}
\end{minipage}
\mbox{\hspace{0.5cm}} 
\begin{minipage}[b]{0.45\linewidth}
\centering
\resizebox{1.0\textwidth}{!}{\includegraphics{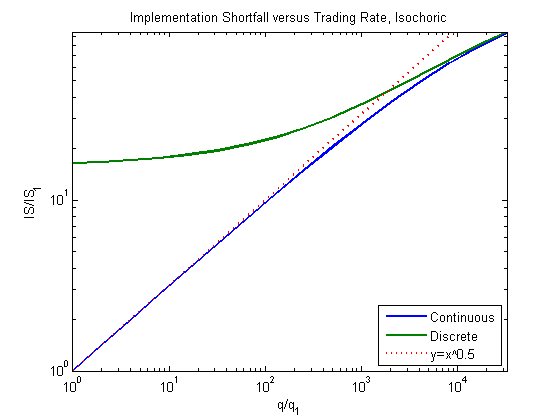} }
\caption{\footnotesize Loglog plot for trading cost versus trading rate from a Figure at left. Discrete (green line) and continuous (blue line) trading}
\label{fig:logImpactQ}
\end{minipage}
\end{figure}
the cost of continuous trading grows linearly with trading rate, time of trade given. The cost of discrete trades is almost constant at the beginning ($16$ trades make $1.25$ times greater impact than the single trade) and then goes asymptotically to the linear function.
Soohun Kim and Dermot Murphy \cite{KM2013} recently reported that ``the average size of an individual transaction has substantially and steadily decreased over time - the average size of a buy or sell order in 1997 is 5,600 shares, while in 2009, it is only 400 shares. However, over this same time period, the average number of consecutive buy or sell transactions has significantly increased, from 2.3 consecutive buy or sell orders in 1997 to 11.9 in 2009''. This process is still continuing. The average size of an order executed
on the NYSE and NASDAQ has declined from 600 shares in 2003 to less than 200 shares in 2010 \cite{BR2013}. It means that unattainable area between green and blue lines on Figure \ref{fig:logImpactT} is shrinking. We used diffusion model for simulation, but in an isochronic regime the concrete form of a reasonable (smooth and monotonic decreasing) kernel is not important . For example, an exponential kernel would generate a similar loglog plot with the flat part even flatter. In empirical studies it would be easy to take those convex curves for slowly growing concave functions, e.g., power or even logarithmic, especially when additional point $(0,0)$ is tacitly assumed. Note, that $R^2$ for model parameters calibrations in this case is less than $1\%$.

Fitting  $VWAP$ data means calibration by millions of transactions made entirely by directional traders. $VWAP$ is just the most popular and easy to interpret algorithm of trading.
In general, \emph{Efficient Trading Hypothesis} stipulates the shape of market impact for real market conditions, naturally weighted and balanced. It is a benchmark for any theoretical models and empirical studies of market impact.



\paragraph{ Square Root Dependence of Market Impact on Trading Volume}

The most popular and generally accepted by practitioners estimation for transaction cost is
\begin{equation}
\label{E:sqrt cost}
\Delta W = C_1 + C_2\cdot \sigma \cdot \sqrt{\frac{Q}{ADV}}
\end{equation}
where $C_1$ and $C_2$ are constants.
Almgren et al. \cite{A2005} analyzed almost $700,000$ ($29,509$ \footnote{Statistical laws are much less applicable to market analysis than to statistical physics.  Compare the size of that sample $(\simeq3\times 10^4)$ with the characteristic sample size in chemistry
- Avagadro's number  $\simeq 6\times10^{23}$ $ mol^{ - 1 }$.  } after filtering) US stock trade orders. They had a maximum of $548$ executions per order with a median around $5$ and the median time around one-half hour. Their empirical studies  result in the similar power law. 
\begin{equation}
\label{E:almgImpact}
h = \eta\cdot\sigma \cdot sgn(q)\cdot \left|\frac{Q}{ADV\cdot T}\right|^{\beta}, \ \beta\approx 0.6
\end{equation}
Uniform rate of trading over a volume time interval $T$ was assumed  $q$, $Q=qT$. Therefore, temporary impact  
\begin{equation}
\label{E:almgImpact_M}
h = \eta\cdot\sigma\cdot sgn(q)\cdot \left|\frac{q}{ADV}\right|^{\beta}
\end{equation}

Conventional wisdom and rigorous data mining both suggest a similar concave dependance of market impact on trading volume.
It seems that the only way to follow this formula is to assume a nonlinear impact function.
And now partial relaxation after discrete trades cannot help. The equation (\ref{E:sqrt cost}) was presented as a generalization of a trading rule of thumb that it costs roughly one day volatility to trade one day's volume. With that volume we are definitely in a continuous trading regime. It seems that our attempt to reconcile linear instant impact and concave impact of  continuous trading failed. To understand what happened, let us look at the more general equations. The cost of a trade at a constant rate is the twice integrated market impact kernel
\begin{equation}
\label{E:costGeneral}
\Delta W = \eta\frac{f(q)q}{Q} \int^T_0{dt \int^t_0{K(\tau)d\tau}}=\eta\frac{f(q)}{T}K_{-2}(T)
\end{equation}
In $GKAC$ model $K(t)=\delta(t)$, $K_{-1}(t)=1$, $K_{-2}(t)=t$ and
\[
\Delta W = \eta \cdot f(q)
\]
The $SKAC$ model temporary impact depends only on trading rate. The specific numerical examples of permanent and temporary impact costs for two large-cap stocks were shown in (Table $3$) of \cite{A2005}. The execution of $10\%$ of $ADV$ shares was completed in $0.5$, $0.2$ and $0.1$ days.
Temporary impact of both stocks purchases follows the law (\ref{E:almgImpact_M}). Those examples directly state that
\begin{equation}
\label{E:almgImpact_M2}
\Delta W \propto \sigma \cdot T^{-\beta}|_{Q=const} \propto \sigma \cdot q^{\beta}|_{Q=const}
\end{equation}
We illustrated an \emph{isochoric} regime of trading on Figures \ref{fig:8tradesQ} and \ref{fig:logImpactQ}.  The value of the exponent $\beta=0.6$ in \cite{A2005} could be explained by a mixed regime in their analysis. \footnote{ In recent paper by J.D Farmer et al. \cite{FGLW2013} the data in \cite{A2005} (among others) are considered as a support of a statement: ``The empirical results strongly support concave dependence on size, whereas the dependence on time is an open question''.}

Grinold \& Kahn \cite{GK2000} give an elegant heuristic derivation of this equation (\ref{E:sqrt cost}), ($chapter \ 16$, $Equation \ 16.4$). They explain that liquidation time is proportional to the size of stock inventory ~($ chapter \ 16, \ Equation \ 16.1$). In our notation
\begin{equation}
\label{E:T clear}
T \propto \frac{Q}{ADV} 
\end{equation}
Substituting into (\ref{E:sqrt cost}), we get
\begin{equation}
\label{E:sqrt cost2}
\Delta W-C_1 \propto \sigma\ \cdot \sqrt{T}
\end{equation}
Again, a more thorough look at this example doesn't confirm concave dependance of temporary market impact on trading rate in an isochronic regime. The assumption under equation (\ref{E:sqrt cost}) was not a fixed time of execution, but an \emph{Efficient Trading Hypothesis} or, in other words, a professional quality trading engine that makes a reasonable choice of a trading rate and keeps it. 
Cost of trade per share in this \emph{isotachic} or \emph{isokinetic} regime $q=const $ is proportional to the square root of time. We don't need a special illustration of this regime of trade - each trajectory on Figures \ref{fig:8tradesT} and \ref{fig:8tradesQ} is isotachic.
Recent empirical research by Bershova and Rakhlin \cite{BR2013} confirms square root dependance on time in an isotachic regime at average. However larger orders in their sample are best approximated by a logarithmic function. The possible explanation of slower growth is in a more optimal than uniform rate execution for larger orders. A frontloading, as shown by equations \ref{E:constprice,E:powerPrice}, can significantly distort the square root law.
Plugging an asymptotic form of a diffusion kernel (\ref{E:solTimeBig}) into a general equation for the implementation shortfall (\ref{E:costGeneral})
\begin{equation}
\label{E:diff cost}
\Delta W = \eta \cdot \frac{q}{T}K_{-2}(T) = \tilde{\eta} \cdot \frac{1}{\sqrt{\kappa}} \cdot q \cdot \sqrt{T}, \ T\gg 1
\end{equation}
Comparing equations (\ref{E:diff cost}) and (\ref{E:sqrt cost}),  we found the meaning of diffusion coefficient $\kappa$ in our `information space'.
This parameter controls the speed of market response and, therefore, controls the volatility of a stock.
\begin{equation}
\label{E:diff coeff}
\sigma \sim  \frac{1}{\sqrt{\kappa}}
\end{equation}

Finally we get a law for all three regimes: if parent order is big enough for continuous rate approximation and doesn't exceed critical value that can crash the market
\begin{equation}
\label{E:ideal_market_law1}
\sigma \cdot \sqrt{T} \cdot  q = \sigma \cdot \frac{Q}{\sqrt{T}} = \sigma \cdot \sqrt{Q} \cdot \sqrt{q}= C \cdot \Delta W
\end{equation}

\emph{For continuous and elastic trading:
isochronic (trading time $T=const$) market impact is linear on trading rate, isochoric (trading volume $Q=const$) market impact is proportional to the square root of trading rate $q$, and isotachic (trading rate $q=const$) market impact is proportional to the square root of trading volume $Q$.}


Noticing that $\sigma \cdot \sqrt{T} $ is a volatility of stock $\sigma^T $ evaluated for period of time $0<t<T$, we can rewrite (\ref{E:ideal_market_law1}) in a more balanced and concise form

\begin{equation}
\label{E:ideal_market_law2}
\frac{\Delta W}{\sigma^T \cdot  q} = \frac{\Delta W \cdot T}{\sigma^T \cdot Q} = C
\end{equation}

\section{Results and discussion}

The diffusion kernel presented in this paper explains many empirical market impact estimations and allows square root \emph{metaorder market impact}  and linear \emph{instantaneous market impact} to coexist. 
We consider rehabilitation of a linear impact model as one of the main results of this paper. 
The importance of linearity was discussed in  B. Toth et al \cite{TL2011} $(TLDLKB)$ \footnote{I want to thank one of the authors, J.-P. Bouchaud, for attracting my attention to this interesting paper }: ``in most systems the response to a small perturbation is linear, i.e., small disturbances lead to small effects''. In contrast, marginal impact for strictly sublinear power functions is singular at $Q=0$. Nevertheless, recognizing that all but linear behavior is highly non-trivial, they admitted it as a real phenomenon and designed the model to justify the anomalous high impact of small trades. Our approach has a lot in common with \cite{TL2011}. As well as in our model of section \ref{S:Diff}, the price distribution in $(TLDLKB)$ \emph{latent order book } satisfies diffusion equation. 
Market impact and all market movements reflect information flow.  We assume that dissemination of information is a diffusion rather than instantaneous process.  It is a requirement of the law that all news and company statements will be available to all market participants at the same time, institutional investors obtain exchange data also at the same time. Obviously it is not a slow and smooth diffusion. What is similar to diffusion is the processing and analysis of publicly available data by investors with different time horizons.
Our model and $(TLDLKB)$ are behavioral approaches, distinct from a pure simulation of physical object $(LOB)$ in the seminal \emph{zero-intelligence} model of \cite{SFGK2003}. $(TLDLKB)$ analyze the  dynamics of market participants' intentions instead of changes in the true order book. The $V$-shaped curve of latent volumes with the tip of $V$ at the current price was assumed to obtain square root impact.

The diffusion model satisfies the no-dynamic-arbitrage principle and can explain empirical results for the all regimes of trading:



\begin{itemize}
  \item  Isochronic (constant time - various volume and rate) market impact cost is a linear function of trading rate.
  \item  Isochoric (constant volume - various time and rate ) market impact cost is a square root function of trading rate.
  \item  Isotachic or isokinetic (constant rate   - various time and volume) market impact cost is a square root function of trading volume.
  \item  Theoretical market impact and real implementation shortfall are not the same. Market impact is decreasing to zero with a decreasing trading rate - implementation shortfall has a minimum.
\end{itemize}

We didn't touch many important aspects of market impact theory in this paper, e.g., relation between temporary and permanent impact, \emph{Efficient market Hypothesis, fair pricing condition} \cite{FGLW2013} and supply-demand balance in general. After the work by Kyle  \cite{aK1985} $(1985)$ it is common to describe market dynamics as a contest of three parties: a single insider who has unique knowledge of `fair' price, noise traders who trade randomly; and market makers who set the prices conditional on trading flow\footnote{$(FGLW)$ \cite{FGLW2013} modified the first agent assuming large number of informed traders with the same long term return prediction.}. But the typical signal of informed long term investors is not interesting for intraday traders because its daily $(information  \ ratio\ll 1)$, market makers try to close all their positions by the end of the day and the retail noisy traders tend to trade following the market. This is another paradox: it is not clear, who is going to take the open positions overnight and why. One can be only sure in \emph{``the subtle nature of `random' price changes''} \cite{BP2004} and the subtle nature of the other market laws and hypotheses.

We don't have the answers to all problems, but we hope to shed some light on them in the next paper by developing optimal trading strategies.

\bibliographystyle{plain}

\appendix
\numberwithin{equation}{section}

 \section{Trading Engine}\label{A:B}

Trading engine is a complex program consisting of pre and post trade analysis, smart order routing, order placement, scheduler and other parts. The structure, architecture and the number of the parts, as well as their  names could be different. We describe the basic principles of \emph{scheduler} and \emph{order placement} modules.

The \emph{scheduler} calculates the optimal trajectory of the parent order(metaorder) trading for each position. The most common orders are \emph{Volume Weighted Average Price (VWAP)} and \emph{Arrival Price (AP)} \cite{SB2005}. If the order is \emph{VWAP}, the trajectory is a \emph{volume curve} - the $U$-shaped plot of the percentage of \emph{Average Daily Volume (ADV)} traded during the day. $AP$ trajectories are optimal in a sense of expected \emph{transaction cost-cost variance} tradeoff.
The output of a \emph{scheduler} module is the number of shares that should be bought/sold in a specific short interval, e.g. ~$5 \  min$.

The \emph{order placement} module splits the number of shares for each time bin into small fractions, typically $100$ shares, and submits that lot to the exchange. Depending on evaluated probability to complete the $5 \ min$ portion of the order in time and client risk aversion the \emph{aggressiveness} parameter is chosen. In the most aggressive mode, all orders are market orders (actually marketable limit or \emph{Immediate Or Cancel (IOC)} ). Otherwise limit orders are being sent inside a spread, on \emph{Best Bid and Offer (BBO)} level or deeper into the \emph{Limit Order Book (LOB)} lot by lot, or as an \emph{Iceberg/Reserve} order \footnote{An Iceberg order allows you to submit an order (generally a large volume order) while publicly disclosing only a portion of the submitted order.}. If the order is not complete at the end of $5 \  min$ interval, our primitive engine will send the rest of it for immediate execution at the last moment. Real \emph{order placement} also requires continuously monitoring the state of the market and calculating high frequency signals like $LOB$ imbalance. Some brokers focus on these signals and predominately use market orders \cite{TE2011}. The threshold of a signal can change based on \emph{aggressiveness} level. To understand why waiting is not always the best policy for getting the best price consider the oversimplified model of the market as a binary tree.
Arrival mid quote price is $S_0$. The return in a unit of time is $\pm 1$ tick. If we send market buy order, we expect the loss
\[
\Delta W=-0.5
\]
If we send limit order on a best bid price we gain a half spread with market step down or have to put our order a tick up
\[
\Delta W = p_-\cdot 0.5 + p_+\cdot (\Delta W - 1), \ p_-=p_+=0.5
\]
and again a binary tree market with equal probabilities of ups $(p_+)$ and downs $(p_-)$ results the same loss
\[
\Delta W = -0.5
\]
Sending market or limit orders in a binary tree market without trading signals makes no difference. This example also shows that a market with long trends is the worst case scenario for market makers and the market with constant mid quote prices is the best.

Directional traders with large parent orders compete with high frequency traders. The former try to hide their intentions and the latter try to detect them. For example, a high frequency trader can make an artificial imbalance to provoke trading engines to send market orders, then immediately cancel her limit order that caused that imbalance. This market manipulation, knowing as $spoofing$, is the high frequency version of the classic `pump-and-dump'. Therefore good trading engines avoid any exact regularities.

\end{document}